\documentclass[journal,article,moreauthors,pdftex]{Definitions/mdpi} 
\preto{\abstractkeywords}{\nolinenumbers}

\firstpage{1} 
\pubvolume{1}
\issuenum{1}
\articlenumber{0}
\pubyear{2021}
\copyrightyear{2020}
\datereceived{}
\dateaccepted{} 
\datepublished{} 
\hreflink{https://doi.org/} 

\usepackage{natbib}
\setcitestyle{numbers,square}

\usepackage{ulem,xcolor,threeparttable}

\Title{Cosmological neutrino N-body simulations of dark matter halo}

\TitleCitation{Cosmological neutrino N-body simulations of dark matter halo}

\Author{Yu Chen $^{1,2,}$\footnote[1]{The first two authors contribute equally to this work.}, Chang-Zhi Lu $^{1,2,}$\footnotemark[1], Juan Li $^{2,5}$, Siqi Liu $^{3,4}$, Tong-Jie Zhang $^{1,2,6}$, and Tingting Zhang$^{7}$}

\AuthorNames{Yu Chen, Chang-Zhi Lu, Juan Li, Siqi Liu, Tong-Jie Zhang and  Tingting Zhang}

\AuthorCitation{Chen, Y.; Lu, C-Z.; Liu, S-Q.; Li, J.; Zhang, T-J.;Zhang,T-T}
\address{%
$^{1}$ \quad Institute for Frontiers in Astronomy and Astrophysics, Beijing Normal University, Beijing 102206, China\\
$^{2}$ \quad Department of Astronomy, Beijing Normal University, Beijing 100875, China\\
$^{3}$ \quad CAS Key Laboratory of Optical Astronomy, National Astronomical Observatories, Beijing 100101, China\\
$^{4}$ \quad University of Chinese Academy of Sciences, Beijing 100049, China\\
$^{5}$ \quad College of Physics and Electronic Engineering, Qilu Normal University, Jinan 250200, China\\
$^{6}$ \quad Institute for Astronomical Science, Dezhou University, Dezhou 253023, China\\
$^{7}$ \quad College of Command and Control Engineering PLA Army Engineering University, Nanjing, China}

\corres{Correspondence: tjzhang@bnu.edu.cn;101101964@seu.edu.cn}

\abstract{The study of massive neutrinos and their interactions is a critical aspect of contemporary cosmology. Recent advances in parallel computation and high-performance computing provide new opportunities for accurately constraining Large-Scale Structures (LSS). In this paper, we introduce the TianNu cosmological N-body simulation during the co-evolution of massive neutrino and cold dark matter components via the CUBEP$^3$M code running on the supercomputer Tianhe-2 and TianNu's connected works.
We start by analyzing $2.537\times10^7$ dark halos from the scientific data of TianNu simulation, and compare their angular momentum with the matched halos from neutrino-free TianZero, revealing a dependence of angular momentum modulus on neutrino injection at scales below 50 Mpc and around 10 Mpc.}

\keyword{Cosmology: theory; large-scale structure of the Universe; methods:numerical; neutrino}

\begin{document}

\section{Introduction}
\subsection{Neutrinos}
Despite being one of the fundamental particles in cosmology, our knowledge of neutrinos' features and interactions is limited.
The absolute mass of neutrinos remains unknown, and its determination is a challenging problem that has important implications for both cosmology and particle physics.

The methods for measuring neutrino mass include three technical approaches: double beta decay, single beta decay, and cosmological observations.
The first method strongly depends on a specific assumption that neutrinos are Majorana-type particles, meaning they are their own antiparticles. The CUORICONO \citep{arnab05prl,2007Double} and recent CUPID-0 \citep{azzol18jpcs} experiments employed this approach.
The second approach is more direct and does not depend on any preset model, only on the relativistic energy-momentum relation. KATRIN experiments, as one of the approach, \citep{2001KATRIN,Th2012Precision} have made great strides, documented in \citep{houdy20jpcs,aker20epjc,aker21prd,aker21jins}.
The third approach involves relic neutrinos, which are the second most abundant particles in the universe after photons. They emerged soon after the Big Bang.
Cosmologists constrain the mass hierarchy of neutrinos by observing their effects on the Cosmic Neutrino Background (CNB) \citep{2013Cosmic,universe8020118}, Cosmic Microwave Background (CMB) \citep{planck20aap5,planck20aap7}, and Large-Scale Structure (LSS) \citep{1999Efficient,2015Planck}.
Many recent works investigate these effects, including haloscope experiments for detecting Cold Dark Matter (CDM) candidates (such as axions and dark photons) \citep{2017Haloscope,milla23prd}, measurements of absolute neutrino mass in \citep{katri22natph} and DUNE \citep{pompa22prl} experiments, and constraints on the CDM parameter(s) \citep{2016Constraints,universe8040201}.
All these attempts reported at least two neutrinos having non-zero masses, with a lower mass limit of $M_{\nu} \equiv \sum M_{\nu,i} \geq 0.05$eV from oscillation experiment \citep{2014ChPhC..38i0001O} and an upper limit of $M_{\nu} \leq 0.12$eV from Planck's CMB observation \citep{planck20aap6}.

Therefore, one can estimate the total mass of neutrinos by observing its cosmological effects and check whether their gravitational effects on the LSS evolution at given $M_{\nu}$ match the current simulations in the non-linear regime, where the previous analytical consideration of structure growth become invalid.
Generally, large neutrino masses will reduce cosmological (<1Mpc) clustering and suppress the power spectrum features of CDM \citep{2014PhRvL.113m1301Z}.

\subsection{N-body simulations}
Particles with mass interact through gravity.
However, no analytical expression exists for the phase space of these particles when their number exceeds 2.
Cosmological numerical evolution requires a large number of particles (beyond one billion), and provides descriptions of their states as an alternative method, which is referred to as the N-body simulation \citep{harder13mn,2013MNRAS.428.3375A}.
It has a well-established history in cosmology since the earliest realization in the 1970s \citep{1970Phonetic,1975Development,1976The,1979N}.

Cosmological N-body simulations can model non-Gaussian effects and provide statistical samples for analyzing the universe at different stages.
The simulation evolves depending on the composition of the universe, which can be determined by the famous Friedmann equation.
Most matter in the universe is massive CDM, so point particle approximation is used to investigate their mutual gravity during evolution.
At present, gravitational interactions are calculated employing particle-mesh (PM), particle-particle (PP), PP plus PM (P$^3$M) and tree algorithms in cosmological simulations with extremely huge scales \citep{2014The}.

Recently, the largest particle-amount N-body simulation, cosmo-$\pi$ \citep{cheng20arx}, was conducted by the supercomputer $\pi$ 2.0 at the weak-scaling efficiency of 95\%, containing $\sim4.39$ trillion CDM particles in a $(3.2/h\ \mathrm{Gpc})^3$ cube with the CUBE code \citep{yu18apjs}.
The Quijote-PNG N-body simulations \citep{coul23apj} were performed to investigate the information on primordial non-Gaussianity that can be obtained beyond today's perturbative regime.
The MillenniumTNG project \citep{heragu22arx} combined the Millennium simulation and IllustrisTNG hydrodynamical galaxy formation model to study matter clustering and halo statistics.
And the SIMBA simulation \citep{elson23newa} found that baryonic matter has tight relations with specific angular momentum $j$ for stars, and the dispersion of stellar $j$-$M$ relation is largely driven by HI content.

\subsection{TianZero and TianNu}
To investigate the effect of massive neutrinos on LSS, massive neutrino particles were incorporated into the TianZero CDM particle simulation, resulting in the TianNu co-evolution simulation.

The TianNu simulation, conducted on the Tianhe-2 supercomputer, is among the largest cosmological N-body simulations to date, surpassing its previous cosmological neutrino simulations by an order of magnitude in scale.
Several of its key indicators remain higher than those of later experiments \citep{cheng20arx}.
The enormous amount of particles, large box size, and high force resolution of TianNu make it an exceptional simulation, facilitating the study of the co-evolution of neutrino-CDM particles and the detailed structure of LSS.

Moreover, the clustering effects of neutrinos and the density fluctuation of CDM are easily affected by the Poisson noise, which may cover the actual signals in the neutrino quantum fluctuations \citep{2013Cosmology}.
Plenty of particles are needed to reduce the Poisson noise in the simulation, so the Tianhe-2 supercomputer can reach sufficient accuracy.

The paper is structured as follows.
In Section 2, we provide an overview of the TianNu simulation.
Section 3 presents the results of the process flow of the TianNu scientific data.
In Section 4 we discuss the simulated impact of neutrinos on CDM halos.
In Section 5, we summarize the conclusions of this paper.

\section{Numerical Methods}
We will review the TianNu simulation first because the data we analyzed in this paper are the CDM-halo part of scientific data from the TianNu simulation, and the simulation was fully performed only once at the maximum computational scale of the Tianhe-2 supercomputer.

\subsection{Parameter preset}
Using the customized parallel N-body simulation code CUBEP$^3$M installed on the Tianhe-2, we perform two simulations inside a cube with a side length of 1200Mpc/h.
One of them is TianZero only containing $6912^3$ CDM particle-groups (every group has $\mathrm{M_{CDMz}}=7\times10^8\mathrm{M_\odot}$ as the mass resolution), and the other is TianNu including $6912^3$ massive CDM particle-groups ($\mathrm{M_{CDM}} _\nu=6.9\times10^8\mathrm{M_\odot}$) and $13824^3$ neutrino ($\mathrm{m}_\nu=0.05$eV) particle-groups (every group has $\mathrm{M}_\nu$ $=3\times10^5\mathrm{M_\odot}$ as the mass resolution to keep $\Omega_\nu/\Omega_\mathrm{M}\approx0.3711\%$).
The only operative difference is that massive neutrinos are injected in TianNu at redshift $z$=5 with $\Omega_M$ fixed, while TianZero is entirely massive-neutrino free.
Moreover, according to \cite{2015Precision}, the initial positions of massive neutrinos were generated by a Gaussian noise map, and their velocity was preset by a linear Zeldovich part and random thermal part.
Two species of massless neutrinos are both added in TianNu and TianZero via the CLASS transfer function \citep{blas11jcap} for the cosmological background, and they had no contribution to the CDM halo mass.
To avoid Poisson noise and sampling variance \citep{inman17prd}, the selected box size is relatively less contaminated by noise at the scale of $0.05<k/(\mathrm{h/Mpc})<0.5$.

The two simulations have the same cosmological model and seed initial condition, imposing a flat Universe by assuming that $\Omega_M + \Omega_{\Lambda} = 1$, where the matter density parameter $\Omega_M = 0.29$ (therefore $\Omega_\nu\approx0.001076$), dark energy density parameter $\Omega_{\Lambda} = 0.71$. Moreover, reduced Hubble constant, initial tilt and mass variances at 8Mpc/h are set to be $h=0.73$, $n_s=0.95$, $\sigma_8=0.84$ \citep{2018ApJ...862...60Q}.
With the Spherical Overdensity algorithm to find halos, we identify CDM halos with at least 100 particles.
Further information regarding the simulations and halo finder is available in \citep{2017NatAs...1E.143Y,2017RAA....17...85E}.

\subsection{Code overview}
We recommend interested readers to look at previous works \citep{2015Precision,2016Increasing} for a detailed analysis of the code structure and technical algorithms relevant to both the pure CDM and neutrino cases. Below, we provide a brief introduction to the CUBEP$^3$M code

The CUBEP$^3$M code gets its name from the cube space, PP force and two levels of the PM algorithm.
The particles are currently stored in a computer-generated cube, while in its predecessor (the PMFAST code \citep{merz05newa}), they were stored in slabs.
PP indicates the interaction between particles is considered in a short-range manner, such as between several adjacent fine grids.
PM means we place particles on meshes of varying roughness and use a coarse grid instead of a group of particles as the minimum unit to calculate the macro and long-range interactions with the gravitational potential \citep{harder13mn} more efficiency.

CUBEP$^3$M is an open-source Fortran code that includes Message Passing Interface (MPI) for inter-node communication and Open Multi-Processing (OpenMP) for intra-node multi-threading \citep{harder13mn}.
Once installed, the codes can be compiled and run from the root directory and its subdirectories.

The code generates a snapshot of the simulated universe by recording the positions and velocities of all particles at a given epoch of cosmic evolution (redshift z) and saving the data in an xv.dat file.
Other expected output from the code includes statistical data on star clusters (halofind.dat), 2D projections of the 3D velocity field (proj\_xy.dat, proj\_yz.dat, proj\_xz.dat), and the power spectrum (cic\_power.dat), among others.
These output files are required when compiling the corresponding code module.

\subsection{Halo data and parameters}
\begin{figure}[htb]
	\widefigure
	\includegraphics[width=13.5cm]{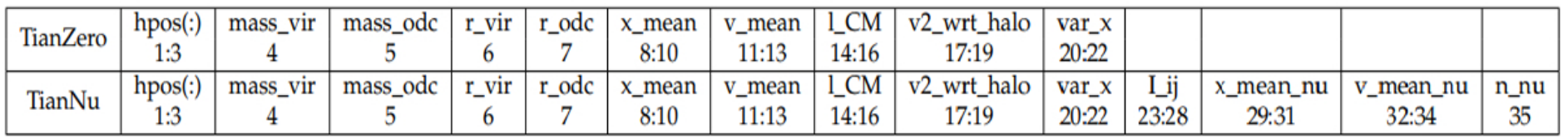}
	\caption{The parameter table for our halo data. The top row is for the TianZero experiment without neutrinos, and the bottom is for the TianNu with neutrinos.
	\label{fig1}}
\end{figure}

Knowing the halo parameter format is essential to understanding CUBEP$^3$M and to conducting our simulation. Therefore, we give our parameter table in Figure \ref{fig1}.

The raw halo data are the snapshot of redshift $z$=0.01.
In Figure \ref{fig1}, we can see that the hpos represents the final halo position (x,y,z), mass\_vir denotes the halo Virial mass, r\_vir is the virial radius of the halo, x\_mean and v\_mean are the position (x,y,z) and velocity (x,y,z) of halo mass center, l\_CM is halo angular momentum modulus (x,y,z), v2\_wrt is the component-squared vector (x$^2$,y$^2$,z$^2$) of the v\_mean, var\_x is the variance (x,y,z) of mass center position.
In the second row, the new x\_mean\_nu and v\_mean\_nu are the position (x,y,z) and velocity (x,y,z) of the halo mass center, while N\_nu is the number density of neutrinos.

Through position and mass data, we can match halos in TianNu and TianZero under specific conditions, such as percent mass variation less than 8\% and mass center position less than 0.08Mpc.
Using angular momenta components data, we can study their moduli changes and direction shifts. With velocities data, we can explore the velocity distribution of the halo mass center.

\section{Data Analysis}
Previous operations provide many halo samples, and have the chance to discover if neutrinos have any significant impact on the halo.

\begin{figure}[htb]
	\centering
	\includegraphics[width=10cm]{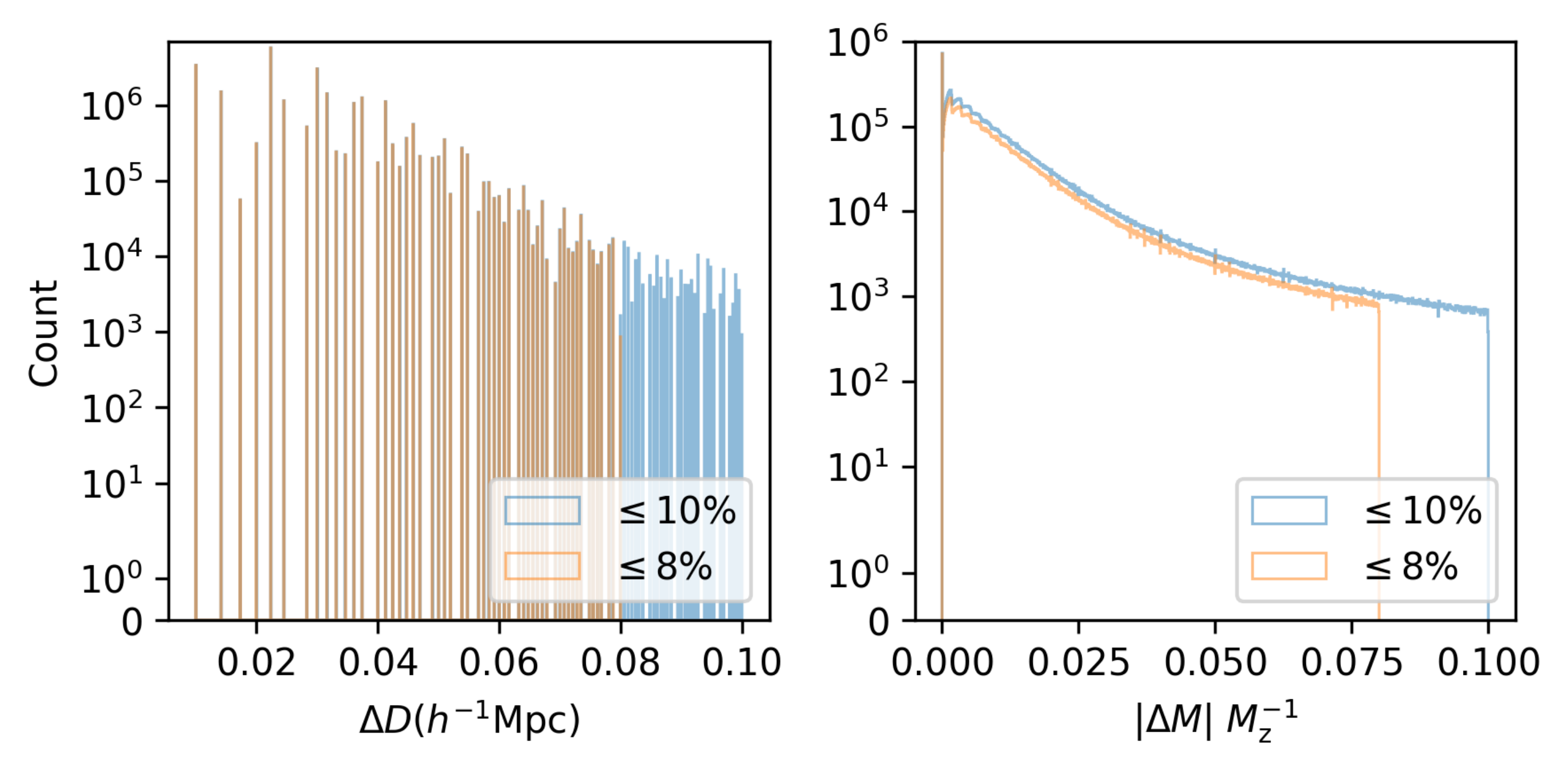}
	\caption{The count of distance and percent mass differences between halo pairs in the left and right panel correspondingly. The M is halo mass, and $\Delta M = \mathrm{abs}(M_\nu-M_\mathrm{z})$. The blue line is the 10\% level halos, and the red line is the 8\% level halos.}
	\label{fig2}
\end{figure}

The study finds $2.560\times10^7$ halos with spherical over-density algorithm \citep{2017NatAs...1E.143Y}. We match the same halos in Figure \ref{fig2} between TianNu and TianZero within a variation tolerance of 10\% (8\%) TianZero mass $M_\mathrm{z}$ and 0.1Mpc (0.08Mpc) center distance respectively. To tighten the match standard, we finally adopt the 8\% level and acquire $2.537\times10^7$ halos, corresponding to the 98.68\% of 10\% level and 97.33\% of whole halos.

Adopting a comparable approach to Bullock et al.\cite{2001MNRAS.321..559B}, we obtain the angular momentum:
\begin{equation}
    \mathbf{J} = \sum_{i=1}^N m_i \mathbf{r}_i \times \mathbf{v}_i.
\end{equation}
After the above process, we obtain halo data, x\_m\_l.dat (x1, x2, x3, M, J1, J2, J3), and calculate the modulus of $\mathbf{J}$, $J=\sqrt{J_1^2+J_2^2+J_3^2}=|\mathbf{J}|$. With the three components of angular momentum, we can compute the variations in its magnitude and direction.

\begin{figure}[htb]
	\widefigure
	\includegraphics[width=10cm]{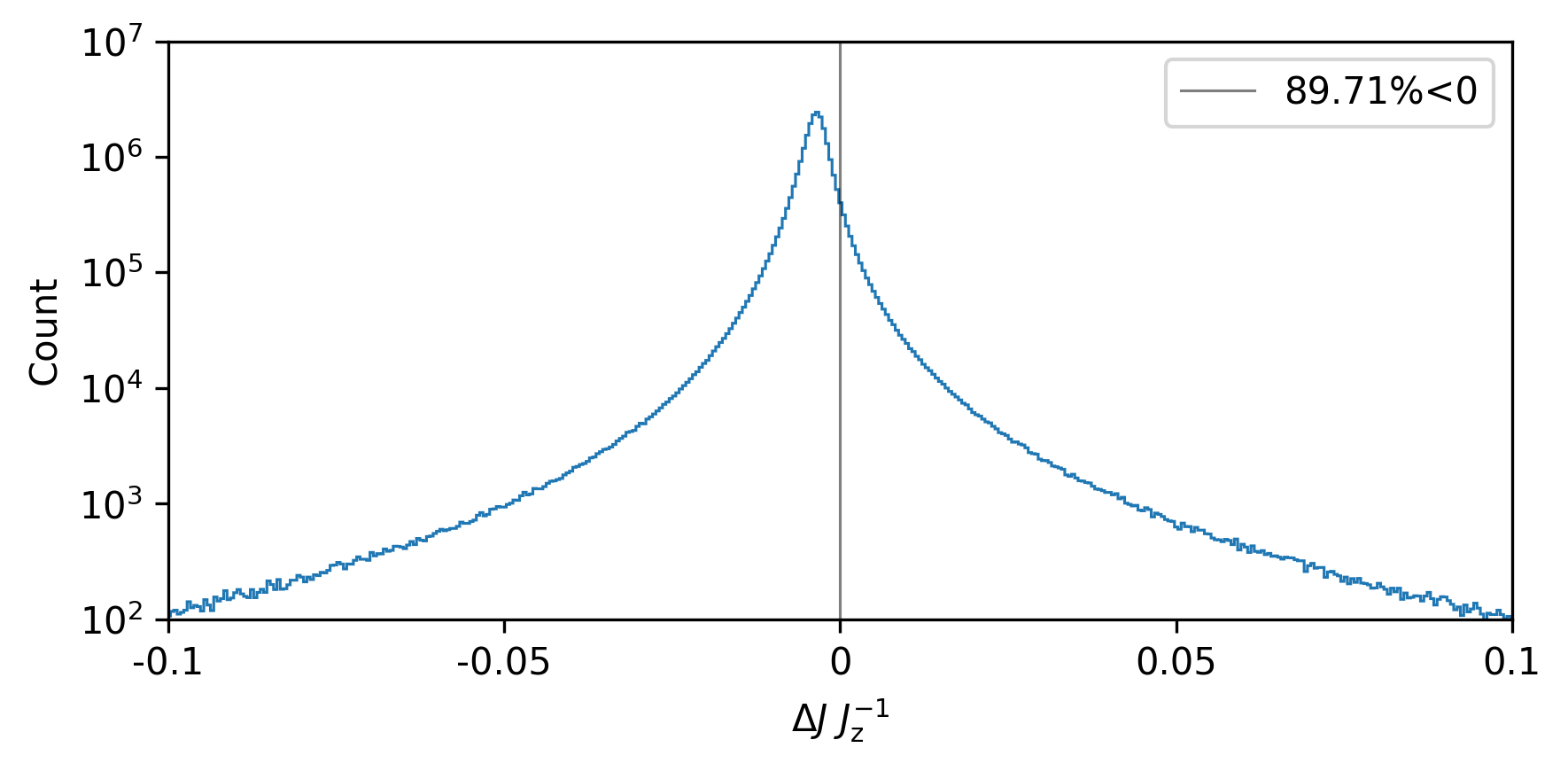}
	\caption{The histogram of the relative modulus variation $\Delta J/J_\mathrm{z}$, where $\Delta J=J_\nu-J_\mathrm{z}$.}
	\label{fig3}
\end{figure}

\begin{figure}[htb]
	\widefigure
	\includegraphics[width=11cm]{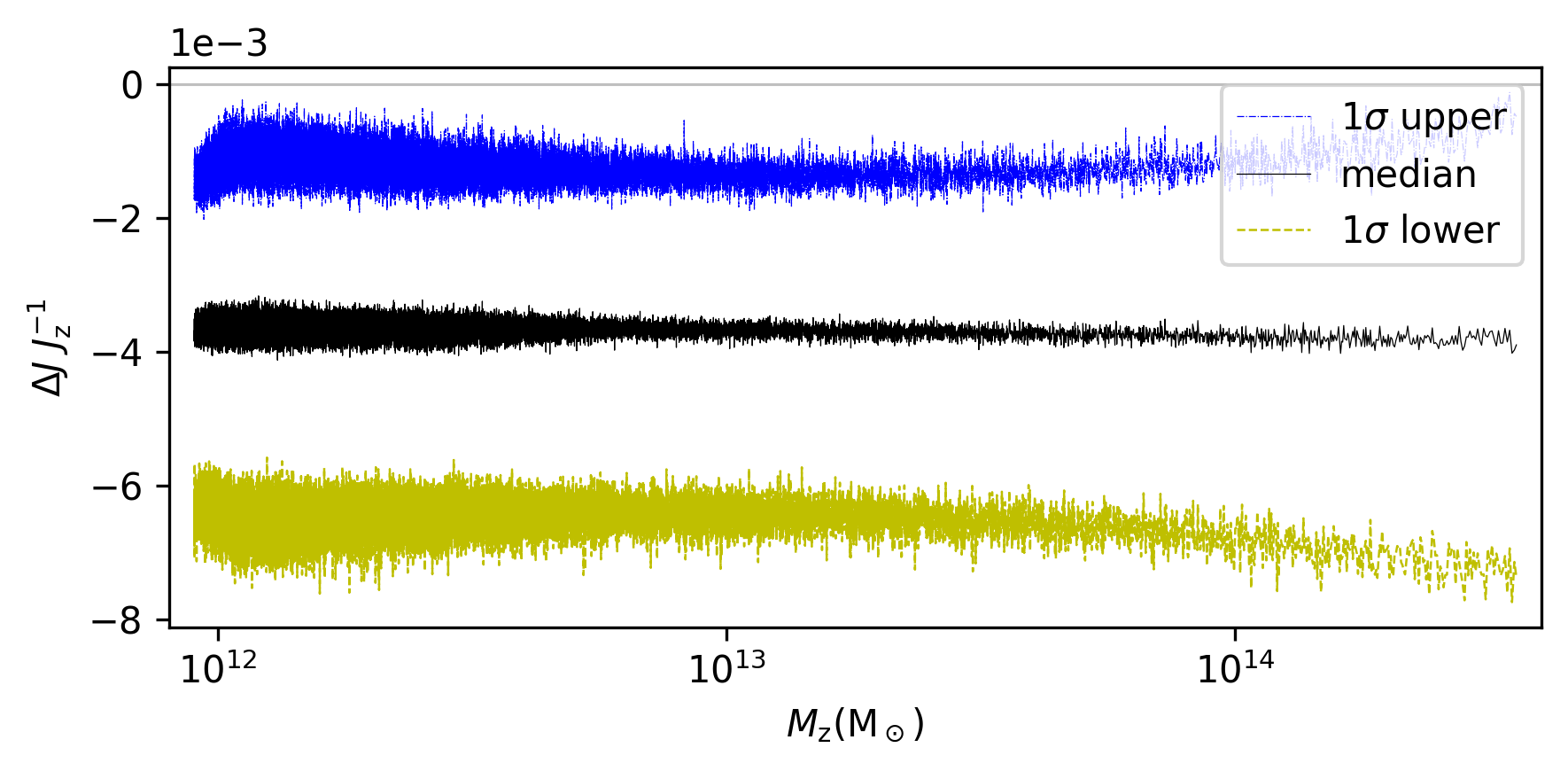}
	\caption{The 1$\sigma$ region of the $\Delta J/J_\mathrm{z}$ median according to $M_\mathrm{z}$. The blue dashed-dotted and yellow dashed lines are the 1$\sigma$ upper and lower limit for the median values (black line).}
	\label{fig3b}
\end{figure}

\begin{figure}[htb]
	\widefigure
	\includegraphics[width=10cm]{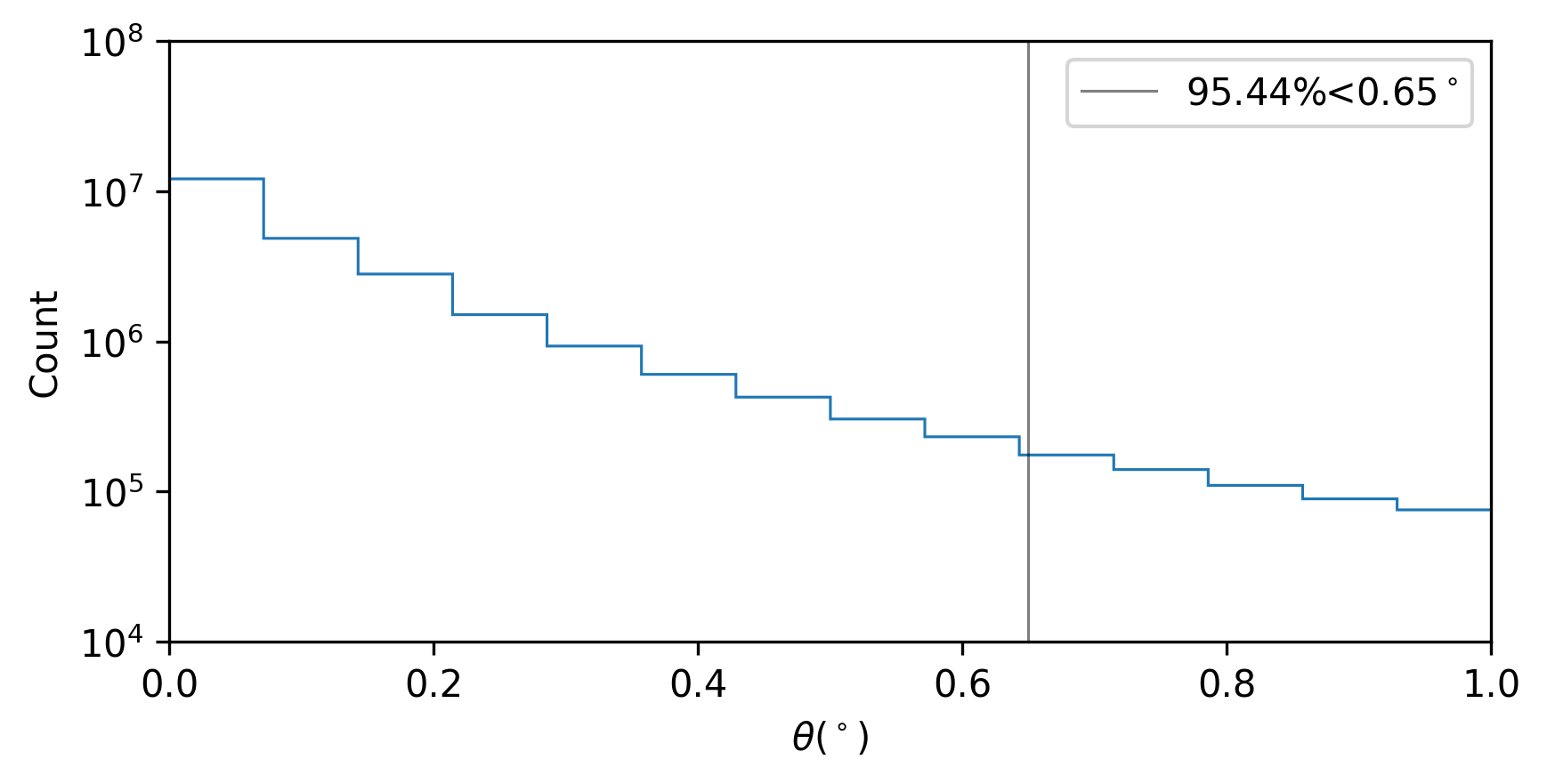}
	\caption{The histogram of the angle of varied \bf{J} direction, $\theta$.}
	\label{fig4}
\end{figure}

\begin{figure}[htb]
	\widefigure
	\includegraphics[width=10cm]{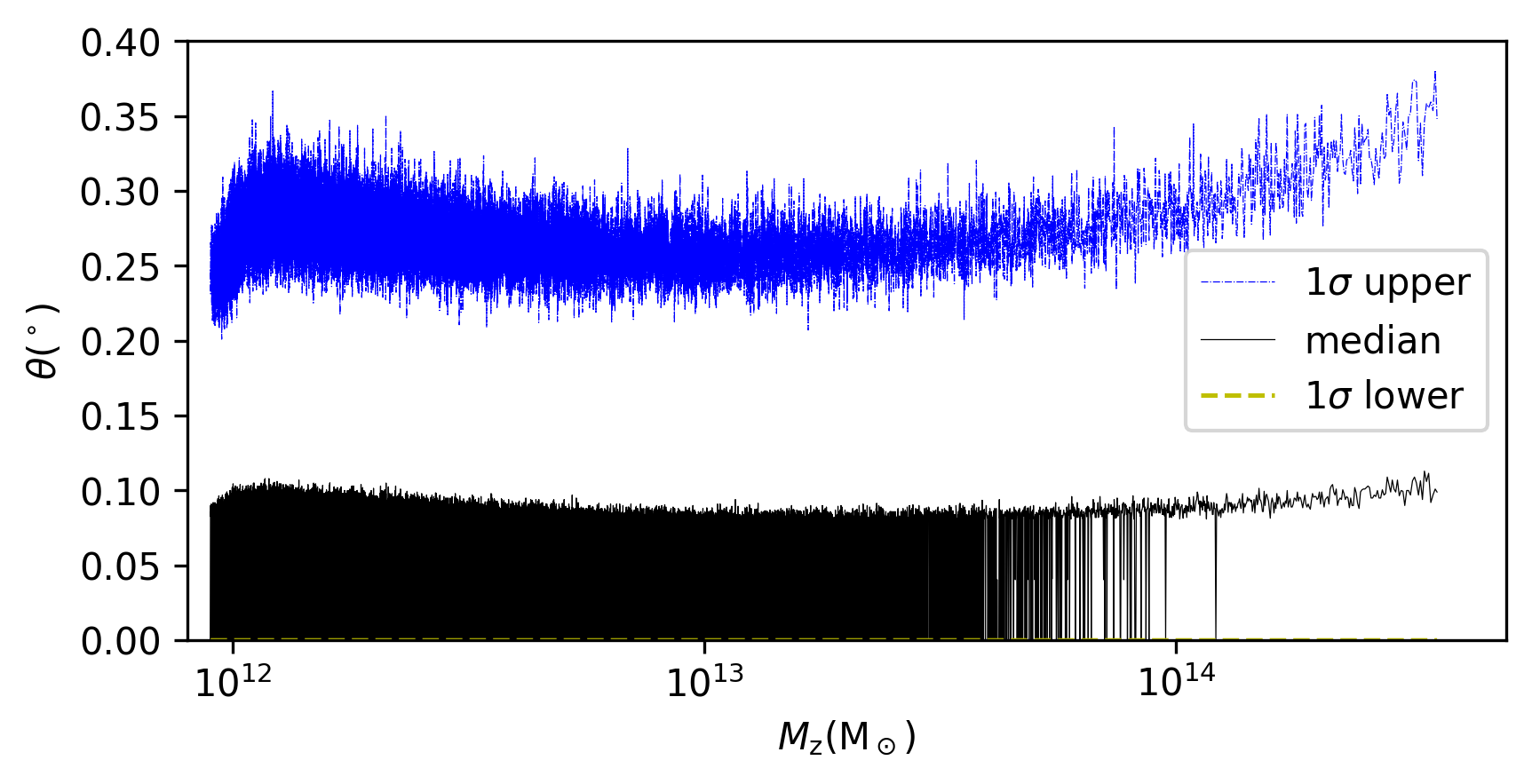}
	\caption{The 1$\sigma$ region of the $\theta$ median according to $M_\mathrm{z}$. The blue dashed-dotted and yellow dashed lines are the 1$\sigma$ upper and lower limit for the median values (black line). The lower limit line is almost invisible because it is almost zero and covered by the black median line.}
	\label{fig4b}
\end{figure}

From Figure \ref{fig3}, we can obtain that the median value of relative modulus variation is -0.3655\% and the most (89.71\%) halos have less angular momenta moduli, indicating that neutrino-injection will reduce most moduli.
We plot the relation between the 1$\sigma$ region of $\Delta J/J_\mathrm{z}$ and $M_\mathrm{z}$ in Figure \ref{fig3b}, where every point (such as median, upper and lower limits) corresponding to one mass is extracted from one thousand dark halos with close masses and different points do not include same halos.
But from Figure \ref{fig4}, the 95.44\% of directions shift less than 0.65 degrees, indicating no significant impact of neutrinos.
And we also plot the relation between the 1$\sigma$ region of $\theta$ and $M_\mathrm{z}$ in Figure \ref{fig4b} with the same 1000-bin statistic in Figure \ref{fig3b}.

Now, we want to explore the universality of neutrinos' effect on halos in different environments.
To expedite the search process, we utilize a two-level division procedure.
Initially, we partition the entire box into adjacent coarse cubes, each with a side-length of 50 Mpc.
Subsequently, we examine every halo within a fine cell (concentric circle) of radius 5 Mpc.
We choose the top 20 densest and 20 sparsest cubes, and select 200 halos from the most compact cell (hereafter referred to as compact halos) and 200 halos from the most sparse cell (hereafter referred to as scarce halos) in each cube.
For clarity purposes, we only use the terms dense or sparse to describe (coarse) cubes and the terms compact or scarce to depict (fine) cells.

\begin{figure}[htb]
	\widefigure
	\includegraphics[width=10cm]{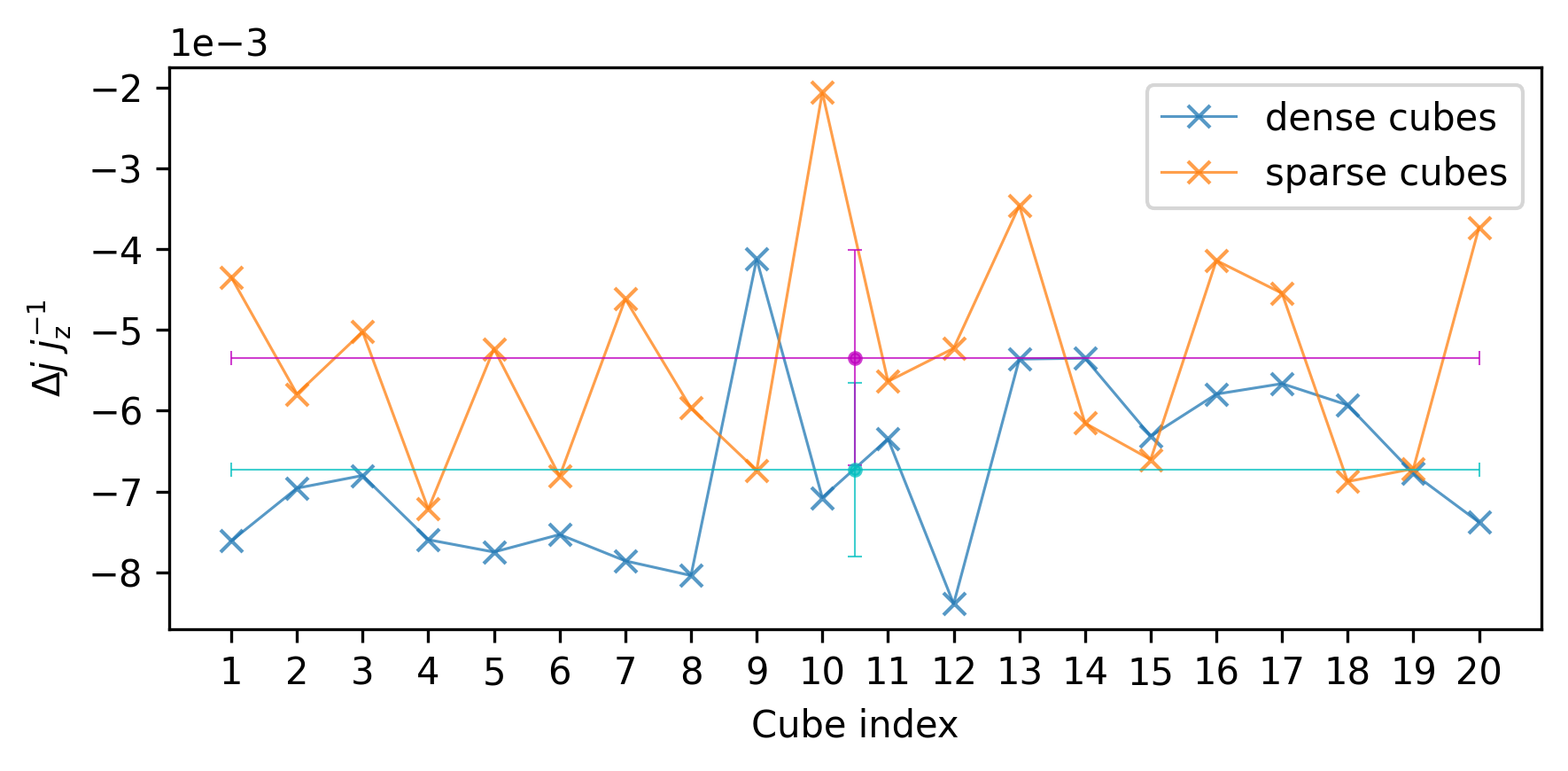}
	\caption{The statistics of mean $\Delta j/j_\mathrm{z}$ of each cube. The blue-cross and red-cross lines are the mean values of dense and sparse cubes. The cyan and magenta vertical error bars are their corresponding standard deviations. On average, every dense cube has 2830 halos with $9.14\times10^{12}\ \mathrm{M_\odot}$, and each sparse one collect 840 halos with $4.37\times10^{12}\ \mathrm{M_\odot}$.}
 	\label{fig5}
\end{figure}

In Figure \ref{fig5}, we plot the averaged $\Delta j/j_\mathrm{z}$ (the variation of mass-reduce angular momentum modulus, and $j=J/\widetilde{M}$, $\widetilde{M}$ is the CDM-particle group number of each halo, namely the mass divided by its mass resolution) for all halos in each selected cube, and present the mean and stand deviation of all averaged $\Delta j/j_\mathrm{z}$.
The mean $\Delta j/j_\mathrm{z}$ of dense cubes is $-0.0067(11)$, while for sparse cubes it is $-0.0053(13)$.
Although their error bars do not touch the other mean values, they overlap with the other bar in most areas.
So their difference exists but is very small.
To provide more information about Figure \ref{fig5}, we list the cubes' statistics obtained from CDM halos in Tables \ref{tab1} and \ref{tab2}.
\begin{table}[htb]
	\centering
	\caption{The statistics of the 20 densest cubes.}
	\label{tab1}
	\begin{threeparttable}
		\begin{tabular*}{\linewidth}{llllllll}
			\toprule
			index & $N_\mathrm{halo,z}$ & $\Delta N_\mathrm{halo}$ $^a$ & $M_\mathrm{total,z}$ & $\overline{M_\mathrm{z}}$ & $\overline{\Delta M/M_\mathrm{z}}$ & $\overline{j_\mathrm{z}}$ & $\overline{\Delta j/j_\mathrm{z}}$ \\
			& & & ($10^{15}\ M_\mathrm{\odot}$) & ($10^{12}\ M_\mathrm{\odot}$) & (\%) & ($\frac{\mathrm{Mpc\ km\ s^{-1}}}{h\ \mathrm{M_r}}$)$^b$ & (\%) \\ 
			\midrule
			1 & 3041 & 2 & 28.53 & 9.37 & -1.41\% & 143.00 &-0.76\% \\
			2 & 2905 & 1 & 27.91 & 9.61 & -1.32\% & 166.92 &-0.70\% \\
			3 & 2678 & 0 & 27.00 & 10.13 & -1.37\% & 166.61 &-0.68\% \\
			4 & 2888 & -1 & 26.93 & 9.33 & -1.23\% & 57.83 &-0.76\% \\
			5 & 2852 & 5 & 26.87 & 9.40 & -2.29\% & 144.55 &-0.78\% \\
			6 & 2870 & 5 & 26.22 & 9.12 & -1.27\% & 165.71 &-0.75\% \\
			7 & 2790 & 1 & 25.91 & 9.28 & -1.43\% & 181.33 &-0.79\% \\
			8 & 2936 & 2 & 25.69 & 8.74 & -1.31\% & 94.63 &-0.80\% \\
			9 & 2644 & 4 & 25.84 & 9.76 & -1.46\% & 75.23 &-0.41\% \\
			10 & 2722 & 3 & 25.67 & 9.42 & -1.26\% & 147.85 &-0.71\% \\
			11 & 2911 & 1 & 25.66 & 8.81 & -1.38\% & 252.50 &-0.64\% \\
			12 & 2623 & 6 & 25.30 & 9.62 & -1.20\% & 92.09 &-0.84\% \\
			13 & 2761 & 1 & 25.24 & 9.14 & -1.45\% & 142.41 &-0.54\% \\
			14 & 2909 & 3 & 25.18 & 8.65 & -1.33\% & 287.76 &-0.53\% \\
			15 & 2952 & 5 & 25.16 & 8.51 & -1.34\% & 164.04 &-0.63\% \\
			16 & 2790 & 5 & 24.89 & 8.91 & -1.54\% & 32.12 &-0.58\% \\
			17 & 2769 & 1 & 24.87 & 8.98 & -1.28\% & 148.10 &-0.57\% \\
			18 & 2791 & 1 & 24.82 & 8.89 & -1.35\% & 84.01 &-0.59\% \\
			19 & 2860 & 1 & 24.76 & 8.65 & -1.29\% & 140.80 &-0.68\% \\
			20 & 2864 & 4 & 24.70 & 8.61 & -1.30\% & 121.08 &-0.74\% \\
			\bottomrule
		\end{tabular*}
		\begin{tablenotes}
			\footnotesize
			\item $^a$ $\Delta N_\mathrm{halo}=N_\mathrm{halo,Nu}-N_\mathrm{halo,Zero}$.
			\item $^b$ $\mathrm{M_r^{-1}}$ is the mass resolution of CDM-particle group.
		\end{tablenotes}
	\end{threeparttable}
\end{table}
\begin{table}[htb]
	\centering
	\caption{The statistics of the 20 sparsest cubes.}
	\label{tab2}
	\begin{threeparttable}
		\begin{tabular*}{\linewidth}{llllllll}
			\toprule 
			index & $N_\mathrm{halo,z}$ & $\Delta N_\mathrm{halo}$ $^a$ & $M_\mathrm{total,z}$ & $\overline{M_\mathrm{z}}$ & $\overline{\Delta M/M_\mathrm{z}}$ & $\overline{j_\mathrm{z}}$ & $\overline{\Delta j/j_\mathrm{z}}$ \\
			& & & ($10^{15}\ M_\mathrm{\odot}$) & ($10^{12}\ M_\mathrm{\odot}$) & (\%) & ($\frac{\mathrm{Mpc\ km\ s^{-1}}}{h\ \mathrm{M_r}}$)$^b$ & (\%) \\
			\midrule
			1 & 743 & -1 & 3.13 & 4.22 & -1.40\% & 77.77 &-0.46\% \\
			2 & 726 & 0 & 3.25 & 4.47 &-1.40\% & 148.00 &-0.58\% \\
			3 & 839 & -1 & 3.28 & 3.92 & -1.43\% & 40.74 &-0.50\% \\
			4 & 855 & 0 & 3.40 & 3.98 & -1.25\% & 99.81 &-0.72\% \\
			5 & 858 & -4 & 3.44 & 4.03 & -0.55\% & 27.99 &-0.52\% \\
			6 & 828 & -1 & 3.53 & 4.27 & -1.14\% & 147.53 &-0.68\% \\
			7 & 831 & -1 & 3.57 & 4.30 & -1.41\% & 32.64 &-0.46\% \\
			8 & 875 & 0 & 3.59 & 4.10 & -1.45\% & 113.56 &-0.60\% \\
			9 & 803 & 1 & 3.63 & 4.51 & -1.19\% & 180.72 &-0.67\% \\
			10 & 813 & -1 & 3.63 & 4.47 & -1.67\% & 95.40 &-0.21\% \\
			11 & 759 & -1 & 3.68 & 4.86 & -1.51\% & 151.03 &-0.56\% \\
			12 & 868 & 0 & 3.83 & 4.41 & -1.33\% & 149.97 &-0.52\% \\
			13 & 822 & -1 & 3.85 & 4.70 & -1.44\% & 40.90 &-0.35\% \\
			14 & 912 & 0 & 3.89 & 4.26 & -1.21\% & 141.78 &-0.61\% \\
			15 & 910 & -1 & 3.97 & 4.37 & -0.003\% & 87.19 &-0.66\% \\
			16 & 858 & -2 & 4.05 & 4.73 & -1.37\% & 150.18 &-0.41\% \\
			17 & 1096 & 0 & 4.12 & 3.76 & -1.23\% & 67.54 &-0.45\% \\
			18 & 856 & 0 & 4.15 & 4.85 & -1.25\% & 64.36 &-0.69\% \\
			19 & 923 & -1 & 4.18 & 4.54 & -1.21\% & 150.18 &-0.67\% \\
			20 & 917 & 0 & 4.19 & 4.57 & -1.30\% & 115.81 &-0.37\% \\
			\bottomrule
		\end{tabular*}
		\begin{tablenotes}
			\footnotesize
			\item $^a$ $\Delta N_\mathrm{halo}=N_\mathrm{halo,Nu}-N_\mathrm{halo,Zero}$.
			\item $^b$ $\mathrm{M_r^{-1}}$ is the mass resolution of CDM-particle group.
		\end{tablenotes}
	\end{threeparttable}
\end{table}

\begin{figure}[htb]
	\widefigure
	\includegraphics[width=12cm]{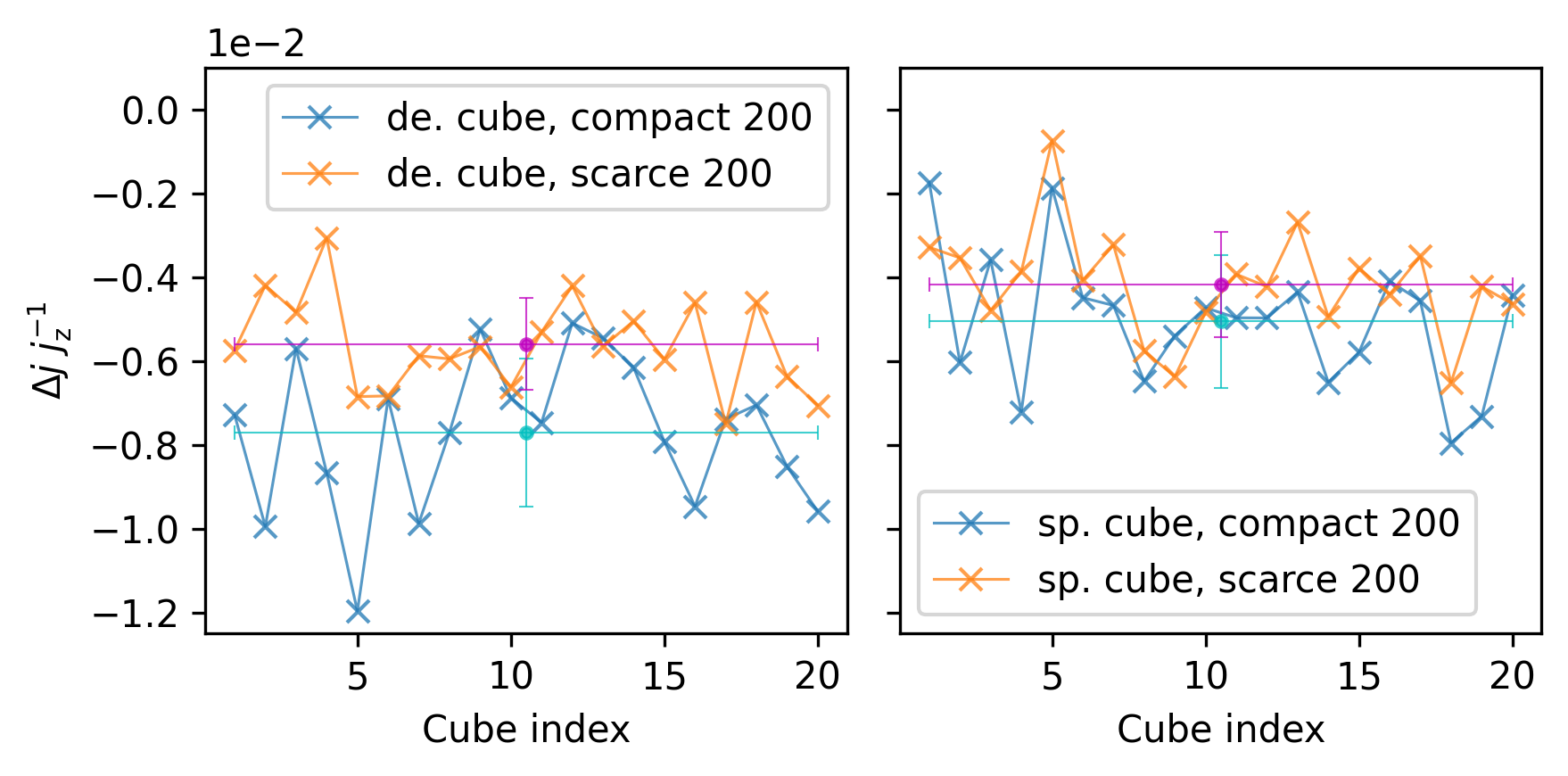}
	\caption{The statistics of mean $\Delta j/j_\mathrm{z}$ of 200 compact and 200 scarce halos in each cube. The left panel shows the case of dense cubes, and the right is for sparse cubes. The blue-cross and red-cross lines are the mean values of 200 compact and 200 scarce halos in each cube, and the cyan and magenta vertical error bars are their corresponding standard deviations. In dense cubes, compact groups averagely have $3.55\times10^{13}\ \mathrm{M_\odot}$ and scarce groups have $1.26\times10^{12}\ \mathrm{M_\odot}$. Similarly, in sparse cubes, compact groups averagely have $8.46\times10^{12}\ \mathrm{M_\odot}$ and scarce groups have $1.28\times10^{12}\ \mathrm{M_\odot}$.}
	\label{fig6}
\end{figure}

In Figure \ref{fig6}, we plot the averaged $\Delta j/j_\mathrm{z}$ for only 200 compact and 200 scarce halos in each selected cube, and handle them as what we do in Figure \ref{fig5}.
The mean $\Delta j/j_\mathrm{z}$ of compact and scarce cells in dense cubes are $-0.0077(17)$ and $-0.0056(11)$, while for the two groups in sparse cubes they are $-0.0051(16)$ and $-0.0042(13)$.
The error bars in the dense cube are more exclusive and distinct.
Moreover, in the compact halo cells in dense and sparse cubes, the two standard deviations of $\Delta j/j_\mathrm{z}$ are always wider than the other two cases in scarce cells, implying different intensity of inner interactions according to their local densities.

\begin{figure}[htb]
	\widefigure
	\includegraphics[width=10cm]{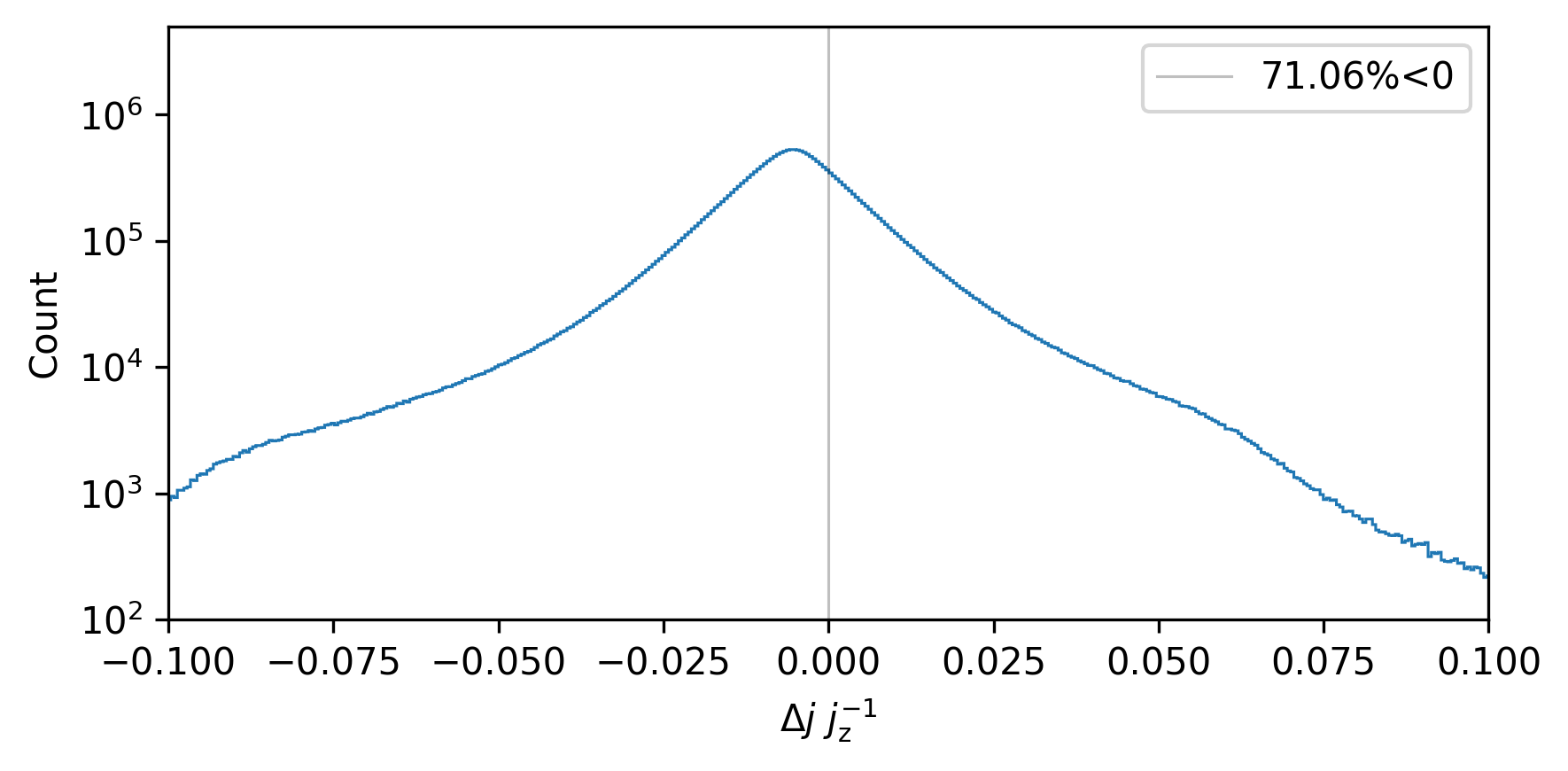}
	\caption{The histogram of the orientation variation $\Delta j/j_\mathrm{z}$.}
	\label{fig7}
\end{figure}

\begin{figure}[htb]
	\widefigure
	\includegraphics[width=10cm]{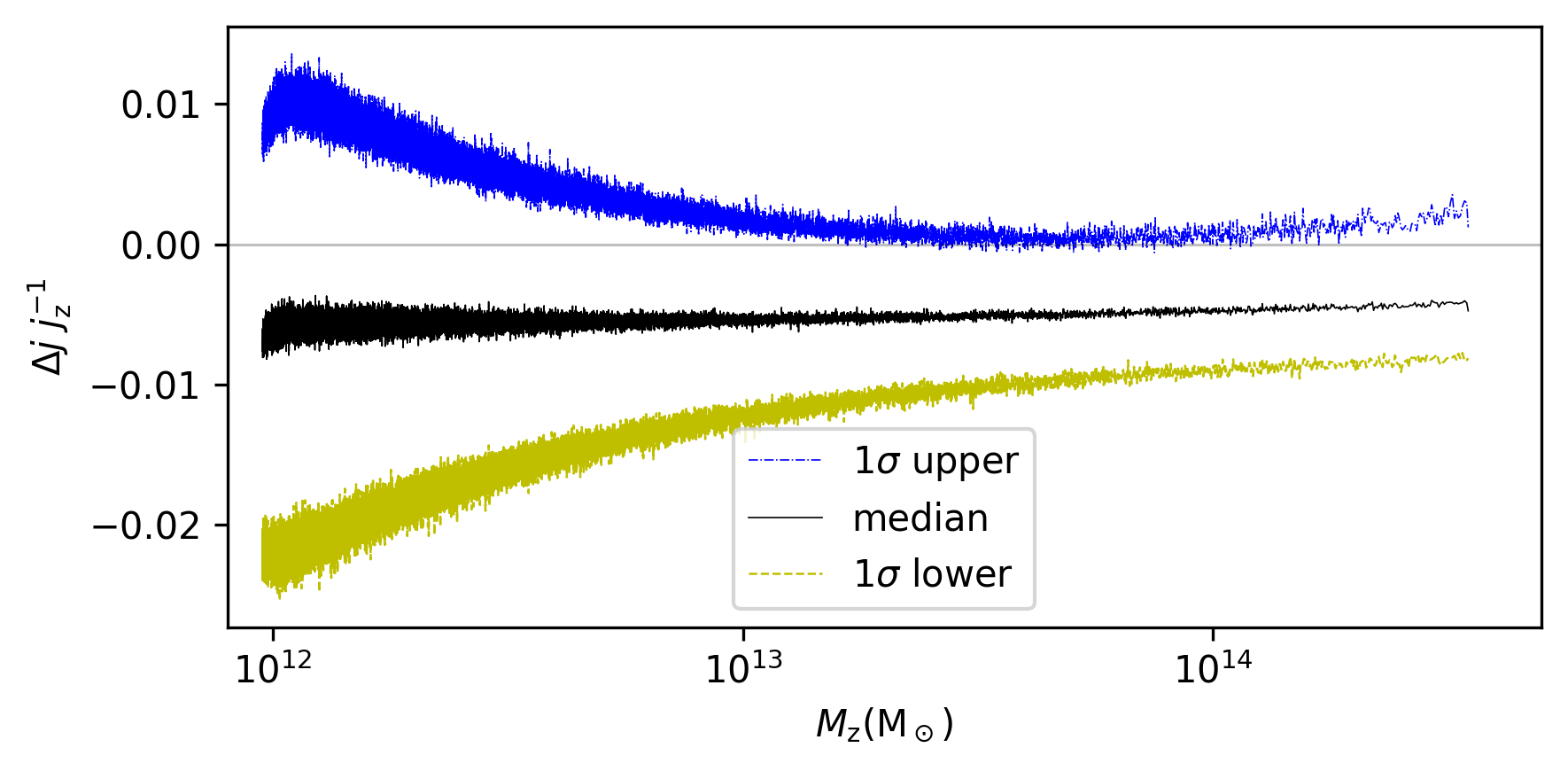}
	\caption{The 1$\sigma$ region of the $\Delta j/j_\mathrm{z}$ median according to $M_\mathrm{z}$. The blue dashed-dotted and yellow dashed lines are the 1$\sigma$ upper and lower limit for the median values (black line).}
	\label{fig7b}
\end{figure}

For comparison, we plot the distribution and median-1$\sigma$ region of $\Delta j/j_\mathrm{z}$ in Figures \ref{fig7} and \ref{fig7b}. 70.06\% samples exhibit a decrease in $\Delta j$ after massive-neutrino injection. The median value of $\Delta j/j_\mathrm{z}$ is -0.0057, which is close to the mean value of 20 sparsest cubes (-0.0056), the 200 scarce halos of the 20 densest cubes (-0.0051).

\section{Discussion}
From Figure \ref{fig2}, the count of distance is discrete.
This comes from relatively limited two decimal places extracted from data, and more digits may alleviate the situation.

The space distribution of LSS is typically irregular, and cannot be adequately represented by simple geometric shapes such as a cube or sphere.
Thus, our local-density statistic of halos is rudimentary and inaccurate (although it is still valuable for simplifying computations).
One possible approach to further studying the LSS distribution is via clustering algorithms, such as K-means, density clustering (e.g., DBSCAN), hierarchical clustering, etc.
However, these new methods face the same challenges, i.e. defining a standard density and addressing the presence of critical halos just outside the analyzed region. Further, incorporating Kernel Density Estimation may provide a unique solution.

Few exceptionally massive halos will contribute importantly to the density statistics, since coarse cubes with more massive halos will be more easily selected. Therefore, many low-mass halos inside the fine cell of each massive halo will be chosen.

The average $\Delta j/j_\mathrm{z}$ for 20 dense and 20 sparse cubes can be compared by examining Figure \ref{fig5}. The error bars for the mean values of these measurements indicate similar interaction intensities on the angular momentum at a cube scale.
Figure \ref{fig6} shows two panels with different average values and standard deviations. The right panel shows error bars that do not exclude the other mean value, unlike its left panel which exhibits a more noticeable difference in values.
But the four cases show some consistency that in higher environment density region, halos incline to have more notable $\Delta j/j_\mathrm{z}$.
Higher density usually comes from some rare but very massive dark halos, which attract more low-mass halos and neutrino groups, naturally resulting in more intensive gravitation.
From our results, this may lead to a potentially more significant loss of total angular momentum modulus.
The two mean $\Delta j/j_\mathrm{z}$ of 200 halos in sparse cubes (the right panel of Figure \ref{fig6}) are both bit lower than the one of all halos in sparse cubes (Figure \ref{fig5}).
Considering the different sample amounts ($\sim$800 and 200) and the mutual coverage of error bars in the 200-halo case, the discrepancy is still acceptable.
In Figure \ref{fig6}, although the average halo mass of scarce halos in dense cubes is one order of magnitude less than that of compact halos in sparse cubes, the former's mean $\Delta j/j_\mathrm{z}$ is double that the latter's, suggesting that the halos' local density environment, rather than their masses, predominantly influences the magnitude of $\Delta j/j_\mathrm{z}$.
Generally, it is more proper to treat the density-dependent $j$-suppression as a statistical rule instead of a certain one.

We select to re-scale $J$ according to its mass resolution (a single particle group) instead of the unit mass $\mathrm{M_\odot}$, because the group is the least simulated object and the unit groups are mass-different in the two simulations. It is more intuitive to compare the performance of these distinct units.
More interestingly, what differences are there among the $j$-distribution of CDM halos, neutrinos and their combination, and are these three distributions in one co-evolution simulation consistent with the one from the single-component simulation? Confined by limited accessible data, these possible discrepancies are beyond our present ability.
Therefore, we advocate more simulations to further explore these questions and examine our conclusions.

The severe non-linear effects in large-scale structure formation make it difficult to provide an accurate theoretical prediction regarding the variance magnitudes of halos' $\Delta J/J_\mathrm{z}$, $\theta$, and $\Delta j/j_\mathrm{z}$ after massive-neutrino injection. For this reason, N-body simulations are being utilized in this area. Our results require more simulations to validate this finding.
However, there have been other studies on the angular momentum modulus of pure CDM halos, aside from the case of CDM-neutrino co-evolution.
Liao et al. \citep{liao17apj} examined the specific angular momentum $j(r,\theta)$ of halos obtained through a high-resolution Bolshoi simulation, and the resulting universal $j(r,\theta)$ profile confirmed the rigid shell model.
Additionally, Li et al. \citep{li22mn} investigated the spin transfer from collisionless CDM to gas during halo formation using the non-radiative cosmological simulation SURFS, providing more detailed results.

Besides our content, several other developments in TianNu warrant a brief review.

Inman et al. \cite{2015Precision} implemented neutrino particles into the cosmology N-body code CUBEP$^3$M  \citep{harder13mn}, which is designed for dark matter simulation as one of its applications, thereby enhancing its functionality. This implementation involved updating the methods used to calculate density and velocity fields.

Emberson et al. \citep{2017RAA....17...85E} optimized the primary TianNu code \citep{2015Precision}, by introducing MPI decomposition for long-range PM force, adding OpenMP parallelism for short-range PM and PP force, and introducing a new data compression method from 24 bytes per particle to 9 bytes. These upgrades ensured the successful execution of the TianNu simulation on the Tianhe-2 supercomputer.

Yu et al. \citep{2017NatAs...1E.143Y} used TianNu simulation to reveal the differential neutrino condensation effect: halo properties like its mass function vary depending on the local neutrino relative abundance (neutrino/CDM), which can be used to infer massive neutrino mass. Massive halos can capture more neutrinos in neutrinos-rich regions than those in neutrino-poor regions.

Inman et al. \cite{inman17prd} proposed that the relative motion between neutrinos and CDM leads to a large-scale dipole on the matter density field which can offer orthogonal constraints to break the degeneracies in the two-point statistics caused by galaxy bias and optical depth in cosmological neutrino mass measurement. The linear response dipole (from relative velocity displacement treated in Fourier space) and N-body dipole (regarded as the result of real space) exhibited a perfect agreement, indicating the possibility of their conception.

Qin et al. \citep{2018ApJ...862...60Q} discovered a local effect from massive neutrinos on the spatial distribution of CDM halos concerning the Delaunay Triangulation (DT) voids with TianNu halo data. Small voids typically locate in regions with more massive neutrinos, making the halos around them more susceptible. The flux of dense massive neutrinos enlarges the void and pushes the neighboring halos away from the void center. Consequently, small voids would become larger than the case without these neutrinos, and large voids will be smaller. This feature impacts the halo two-point correlation at $\sim$1Mpc/h and notably disrupts the function of numerical DT voids. It will play a vital role in current or future galaxy surveys on measurable neutrino impacts.

Apart from TianNu simulation and CUBEP$^3$M code, Yu et al. \citep{2019PhRvD..99l3532Y} utilized a new information-optimized CUBE code \citep{yu18apjs}, which greatly optimized the memory utilization by reducing the memory space per particle from 24 bytes to 6 bytes, and proposed a parity-odd gravitational effect from massive neutrinos, which uniquely torques the direction of the combined angular momentum field of the galaxies and halos. More importantly, this torque detection is free of contamination from linear perturbation and easily separated from other non-gravitational influences. A 21cm HI survey (like HSHS \cite{peter06arx}) aiming at galaxies with $z\approx1$ and $M_\nu$=0.05eV, can reach a 5-$\sigma$ significance of neutrino mass detection.

Furthermore, Kohji et al. \citep{Kohji21arx} employed a combination of codes on the Fugaku supercomputer rather than a single n-body co-evolution program, in their simulation of the first Vlasov simulation of relic massive neutrinos and N-body simulation of CDM. The integration of the Vlasov equation within the 6-dimensional phase space was accomplished using a high-order Vlasov solver enhanced by a series of numerical methods and special instructions for specific processors. The simulation, which involved 400 trillion Vlasov grids and 330 billion-body computations, consistently reproduced the accurate nonlinear neutrino dynamics. Compared to TianNu, the simulation required only about 2 hours for time-to-solution with higher velocity resolution, earlier massive-neutrino injection, and an equivalent realization level in contrast to TianNu's 52 hours.

\section{Conclusions}
This paper presents the TianNu simulation and its key parameters, including halo data and associated scientific outcomes.
Currently, a growing number of scholars are using CUBEP$^3$M to investigate various aspects of neutrinos, cosmic tides and magnetic fields.
We aim to leverage the existing data and analysis framework for future endeavors.

By incorporating massive neutrinos into the CUBEP$^3$M N-body code, we generate fruitful scientific outcomes.
These outcomes, particularly concerning the absolute neutrino mass and hierarchical open issues on the large-scale structure effects in our Universe, expand our understanding.

Our findings suggest that 89.71\% of halo samples display a decreasing trend in the angular momentum modulus when massive neutrinos are added.
Halos located in denser local environments with a diameter of approximately 10Mpc exhibit a more prominent descending trend, implying that halos situated within cosmic filaments display smaller moduli with the presence of massive neutrinos.
Meanwhile, the mean value of the relative variation of mass-reduced modulus ($\Delta j/j_\mathrm{z}$) for 200 compact halos in different dense cubes has a large standard deviation, suggesting that the decreasing effect from massive neutrinos is more statistical, facing the complicated and intensive interplay in over-dense regions.
At last, we review some remarkable work related to TianNu.

We hope this paper will be beneficial for future N-body simulations when browsing the TianNu project, and even to further improve the code.

\acknowledgments{We thank the anonymous referee for the instructive comments and careful discussions.
This work was supported by the National Science Foundation of China (Grants No. 11929301, 61802428).}

\vspace{6pt}

\end{paracol}

\reftitle{References}

\externalbibliography{yes}
\bibliography{neutrino}

\end{document}